\documentstyle[12pt]{article}
\begin{document}
\newcommand{\be}{\begin{eqnarray}}
\newcommand{\ee}{\end{eqnarray}}
\def\ni{\noindent}
\def\ul{\underline}
\def\in{\indent}
\def\lin{\in\in\in}
\def\sss{\scriptscriptstyle}
\pagestyle{empty}
\setlength{\baselineskip}{12pt}
\ul{Submitted to {\it Phys. Lett. B}} \hfill \fbox{{\small NTGMI--94--4
Preprint}}
\setlength{\baselineskip}{18pt}
\vskip 3.5cm

\begin{center}

{\Large{\bf Neutrino Capture Cross Sections for $^{\sss 40}$Ar}}\\
{\Large{\bf and $\beta$-decay of $^{\sss 40}$Ti }}

\vskip 0.6cm
{\sf W. E. Ormand}$\,^{1,\dagger}$\ {\sf P. M. Pizzochero}$\,^{2}$, \ {\sf
P. F. Bortignon}$\,^{2}$ \ and \ {\sf R. A. Broglia}$\,^{2,3}$

\vskip 0.5cm

$^1$ {\it Physics Department 161-33 and W. K. Kellogg Radiation Laboratory,
\\ California Institute of Technology, Pasadena, CA 91125, U.S.A.}

\vskip 0.2cm

$^2$ {\it Dipartimento di Fisica, Universit\'a degli Studi di Milano and \\
Istituto Nazionale di Fisica Nucleare, Sezione di Milano,\\ Via Celoria 16,
20133 Milano, Italy}

\vskip 0.2cm

$^3$ {\it The Niels Bohr Institute, University of Copenhagen,\\ Copenhagen,
Denmark}

\end{center}
\vskip 0.4cm

\vfill

\centerline{\bf ABSTRACT}
\vskip 0.3cm

Shell-model calculations of solar neutrino absorption cross sections for
$^{40}$Ar, the proposed component of the ICARUS detector, are presented. It
is found that low-lying Gamow-Teller transitions lead to a significant
enhancement of the absorption rate over that expected from the Fermi
transition between the isobaric analog states, leading to an overall
absorption rate of 6.7 SNU. We also note that the pertinent Gamow-Teller
transitions in $^{\sss 40}$Ar are experimentally accessible from the
$\beta$-decay of the mirror nucleus $^{\sss 40}$Ti. Predictions for the
branching ratios to states in $^{\sss 40}$Sc are presented, and the
theoretical halflife of 53~ms is found to be in good agreement with the
experimental value of $56^{+18}_{-12}$~ms.

\vskip 0.4cm
\ni
{\it Subject headings:} Nuclear Structure -- Solar Neutrino Spectroscopy.
\vskip 0.5cm
\vfill
\noindent $^\dagger$ {\footnotesize Present address: Department of Physics,
401 Nielson Hall, University of Tennessee, Knoxville, TN 37996-1200 and Oak
Ridge National Laboratory, MS-6373 Building 2008, Oak Ridge, TN 37931-6373}
\newpage
\pagestyle{plain}
\setlength{\baselineskip}{23pt}

The basic feature of solar neutrino astronomy is to provide a tool that
permits the direct examination of processes that occur in the interior of
the Sun. Neutrinos interact only via the weak interaction, and have
essentially an infinite mean-free path in a normal stellar medium.
Therefore, neutrinos observed on Earth probe solar processes that are
occurring in the present, as opposed to photons, which emerge after
$\approx 10^4$~yr.

Considerable interest has been generated by four solar neutrino
experiments~\cite{1,2,3,4} that yield results that  are different from
those expected from the combined predictions of the standard solar model
and the standard  electroweak theory with zero neutrino masses. In essence,
there seems to be a significant suppression in the neutrino flux over that
predicted from the combined standard models. Two primary questions then
remain: (1) do these experiments require new physics beyond the standard
model for electroweak interactions, or (2) is  the standard solar model  at
fault? A recent study~\cite{5} of over 1000 precise solar models concludes
that it is not possible to simultaneously describe all four experiments
within the framework of standard solar models, and suggests that physics
beyond the standard electroweak theory is required. However, before
definitive conclusions on the presence of new physics can be reached,
further experimental verification is warranted. Indeed, three of the four
experiments were sensitive to different parts of the solar neutrino
spectrum, and it remains to be decided if the fluxes from all neutrino
sources are suppressed, or if some mechanism suppresses higher-energy
neutrinos, such as those from $^{\sss 8}$B, so that the Ga experiments
(GALLEX~\cite{3} and SAGE~\cite{4}) may be interpreted as detecting the
full flux from the pp-chain.

One proposal to examine higher energy solar neutrinos is the Imaging of
Cosmic and Rare Underground Signals (ICARUS)~\cite{6,6b} experiment in
which liquid argon (primarily $^{\sss 40}$Ar) is the detector medium. As
applied to solar neutrinos, ICARUS expects to measure the flux of solar
$^{\sss 8}$B neutrinos by both elastic scattering and absorption. In
addition, by measuring the ratio between absorption events and elastic
scattering events, which can be induced by all neutrino types, it is
possible to deduce the probability of oscillations of electron neutrinos
into $\mu $  and $\tau $ neutrinos independently of solar models~\cite{6b}.
Clearly, accurate knowledge of the neutrino absorption cross section is
needed. In the original proposal~\cite{6,7}, the feasibility of $^{\sss
40}$Ar as a detector medium was assessed by assuming that neutrino
absorption is dominated by Fermi transitions to the isobaric analog state
(IAS) in $^{\sss 40}$K. With this assumption (and the imposition of a 5~MeV
cutoff on the minimum energy of the emitted electron), the solar neutrino
absorption cross section is $\Sigma_{\rm tot}\!\mid_{\sss {\rm IAS}} = 3.8
\times 10^{-43}$~cm$^2$, corresponding to a capture rate on  $^{\sss 40}$Ar
nuclei $ {\cal R}(^{\sss 40}{\rm Ar})\!\mid_{\sss {\rm IAS}} = 2.2$~SNU.
 From the compilation of nuclear levels~\cite{Endt}, however, we note that
there are at least six $J^\pi=1^+$, $T=1$ levels in $^{\sss 40}$K with
excitation energies  lower than the isobaric analog state. Given the
dependence in the neutrino absorption cross section on the square of the
energy of the emitted electron, these low-lying levels may contribute
significantly to the overall neutrino absorption cross section.

Another branch of nuclear physics of great interest is the study of exotic
nuclei, in which radioactive beams are increasingly used to study the
properties of nuclei along the proton and neutron drip lines. Of particular
interest to the ICARUS experiment is the $\beta$-decay of $^{\sss 40}$Ti,
which is the mirror of $^{\sss 40}$Ar. Since isospin is a nearly conserved
quantity (to within a few percent), the Gamow-Teller matrix elements
pertinent to $^{\sss 40}$Ar neutrino absorption can be determined
experimentally from the branching ratios of the $\beta$-decay of $^{\sss
40}$Ti to levels in $^{\sss 40}$Sc (the mirror of $^{\sss 40}$K). In
addition, because of the large Q-value ($Q_{\beta} = 11.737$~MeV), all the
transitions of interest in $^{\sss 40}$Ar can be studied directly. The main
difficulties that need to be overcome in this experiment are the short
lifetime, measured to be $\tau_{1/2} = 56^{+18}_{-12}$~ms~\cite{8}, and the
fact that the $\beta$-decay occurs to states that are proton unbound in
$^{\sss 40}$Sc. The current experimental situation is that branching ratios
have only been measured in which the decay proceeds via proton emission to
the ground state of $^{\sss 39}$Ca. Already at this stage, however, it can
be deduced that Gamow-Teller transitions play an important role as the
expected lifetime for $^{\sss 40}$Ti in the limit of a pure Fermi
transition is 157 ms. In addition, $20\pm 4$\% of all decays are found
experimentally~\cite{8} to proceeded via $\beta$-decay to the second
$J^\pi=1^+$ state in $^{\sss 40}$Sc (at 2.7~MeV) followed by proton
emission to the ground state of $^{\sss 39}$Ca.

In this paper, we present the results of a shell-model calculation for the
Gamow-Teller transitions between the $J^\pi=0^+$, $T=2$ ground state of
$^{\sss 40}$Ar(Ti) to $J^\pi=1^+$, $T=1$ states in $^{\sss 40}$K(Sc). We
find that these low-lying Gamow-Teller transitions significantly enhance
the solar neutrino absorption cross section, increasing the cross section
over that from the isobaric analog transition by nearly a factor of three,
namely up to $\Sigma_{\rm tot } = 11.5 \times 10^{-43}$~cm$^2$ and
correspondingly $ {\cal R}(^{\sss 40}{\rm Ar}) = 6.67$~SNU. For the purpose
of comparison with future experiments, the branching ratios for the
$\beta$-decay of $^{\sss 40}$Ti are also presented. We note that the
deduced half-life of 53~ms is in very good agreement with the experimental
value of $56^{+18}_{-12}$~ms.

Nuclei lying across major shells, such as $^{\sss 40}$Ar, pose a serious
challenge to the shell model approach as two major oscillator shells must
be included in the calculation, e.g. the $0{\rm d}_{5/2}1{\rm s}_{1/2}0{\rm
d}_{3/2}$ (sd) and the $0{\rm f}_{7/2}1{\rm p}_{3/2}1{\rm p}_{1/2}0{\rm
f}_{5/2}$ (fp) shells. Perhaps the most significant problem in performing
calculations within this model space is the large number of configuration
accessible. Indeed, it is not possible to carry out an unrestricted
calculation for anything but the lightest nucleus in this model space. For
this reason, a truncation on the model space must be imposed. Towards this
end, we impose an $n\hbar\omega$ truncation, in which $n\hbar\omega$
denotes the excitation of $n$ particles outside of the lower oscillator
shell (in this case the sd-shell). A severe limitation, even within this
approach, is the so-called ``$n\hbar\omega$ truncation catastrophe''. Since
the expansion of the shell-model wave functions in an $n\hbar\omega$ model
space converges slowly and the dimensions increase rapidly with $n$, one is
often forced to use an effective interaction developed for use in the
$0\hbar\omega$ space, which accounts for the gross properties of the rest
of the series in an approximate way. Because of the slow convergence in
$n$, the $n\hbar\omega$ catastrophe occurs even for $n=2$. In essence, for
mixed $(0+2)\hbar\omega$ calculations, the very strong interaction between
the low-lying $0\hbar\omega$ and $2\hbar\omega$ states with similar
symmetries causes the $0\hbar\omega$ states to be pushed considerably lower
in energy, leaving an unrealistic gap in energy. In addition, not only are
the binding energies grossly in error, but also the mixing between the 0
and 2$\hbar\omega$ states is incorrect because it is dependent on the
perturbed energies. Naturally, if dimensional considerations are not a
concern, then one could at least partially solve the problem of the $2\hbar
\omega$ states with the inclusion of $4\hbar\omega$ states.

Because of the problems inherent in an $n\hbar\omega$ truncation, we follow
the example of ref.~\cite{9} and diagonalize the $0\hbar\omega$ and
$2\hbar\omega$ spaces separately. In this light, the low-lying
positive-parity $T=1$ and $T=2$ states in $A=40$ are purely $2\hbar\omega$
states, while the ground state of $^{\sss 40}$Ca is the only $0\hbar\omega$
state. In addition, since the $2\hbar\omega$ states are not constructed by
the excitation of particles of the same type out of the sd-shell, there are
no spurious center-of-mass states in the calculation. The Hamiltonian used
here is that of ref.~\cite{9}, which consists of the Wildenthal matrix
elements for the sd-shell~\cite{10}, McGory's (0f,1p) shell Hamiltonian for
the fp-shell matrix elements~\cite{11}, and a modification of the
Millener-Kurath potential for the cross-shell interaction~\cite{12}. For 25
nuclei with $Z=13-20$ this interaction reproduced the ground-state binding
energies with an rms deviation of 305~keV. In this work, the wave functions
were computed using the shell-model program OXBASH~\cite{13} on VAX-4000/60
computers.

After diagonalizing the resulting Hamiltonian, and obtaining the wave
functions, we calculate the neutrino capture cross section for the reaction
\be
\nu \ + \ ^{\sss 40}{\rm Ar} \ \longrightarrow \ ^{\sss 40}{\rm K} \ + \
e^- \ , \label{c}
\ee
for all states in $^{\sss 40}$K with excitation energy up to the
particle-decay threshold at 7.58~MeV~\cite{Endt}. The cross section for
absorbing a neutrino with energy $E_\nu$ from the ground state of $^{\sss
40}$Ar to the $i^{th}$ excited state in $^{\sss 40}$K is given by
\be
\sigma_i(E_{\nu}) \ &=& \ \frac{G^2_{\sss V}}{\pi c^3 \hbar^4} \ \mid {\cal
M}_{o \rightarrow i} \mid^2 E_e^i \ k_e^i \ F(Z,E_e^i).
\ee
Here $F(Z,E_e^i)$ is the Fermi function associated with Coulomb correction
factor appropriate for the charge density of the daughter nucleus, $G_{\sss
V}$ is the vector coupling constant for nuclear weak processes, while
$k^i_e$ and $E^i_e$ are the electron momentum and energy, respectively.
They are given by
\be
E_e^i \ &=& \ E_{\nu} \ - \ Q_i \ + \ m_e c^2 \nonumber \\ (k_e^i)^2 \ &=&
\ (E_e^i)^2 - m_e^2 c^4,
\ee
with the Q--value being determined by the difference in binding energy for
the initial and final states,
\be
Q_i \ = \ E_i \ - \ E_o \ + \ m_e c^2.
\ee
The square of the transition matrix element $\mid {\cal M}_{o \rightarrow
i} \mid^2$ is written as
\be
\mid {\cal M}_{o \rightarrow i} \mid^2 \ = \ \left[ {\rm B(F)}_{o
\rightarrow i} \ + \  {\rm B(GT)}_{o \rightarrow i} \right], \label{a}
\ee
where, in the long--wavelength limit, the Fermi and Gamow--Teller reduced
transition probabilities are given by
\be
{\rm B(F)}_{o \rightarrow i} \ &=& \ \frac{1}{2 J_o + 1} \ \left| \langle
J_i \mid \mid t_{\pm} \mid \mid J_o \rangle \right|^2 \nonumber \\ &=& \
\left[ T_o (T_i + 1) - T_{z\,o} T_{z\,i} \right] \ (1-\delta_{\sss C})
\delta_{o,i},
\ee
and
\be
{\rm B(GT)}_{o \rightarrow i} \ &=& \ \frac{1}{2 J_o + 1} \ \left(
\frac{g_{\sss A}}{g_{\sss V}}\right)^2 \ \left| \langle J_i \mid \mid
(\vec{\sigma} \ t_{\pm})_{\rm eff} \mid \mid J_o \rangle \right|^2 .
\label{b}
\ee
The quantity $g_{\sss A}/g_{\sss V} = 1.2606 \pm 0.0075$ \cite{11a} is the
ratio of the axial and vector weak coupling constants and $\delta_{\sss C}$
is the isospin-mixing correction to the Fermi matrix element, which is of
the order 0.2-0.4\%~\cite{11b}, and can be neglected. In keeping with the
observation that experimental B(GT) values are quenched relative to
theoretical estimates, we have multiplied the free-nucleon Gamow-Teller
operators by 0.775~\cite{11c}.

The total absorption cross section for solar neutrinos is obtained by
folding the cross section defined in Eq.(2) with the normalized solar
neutrino flux $\Phi(E_\nu)$, and summing contributions due to excited
states, i.e.
\be
\Sigma_{\rm tot} = \sum_i \Sigma_i = \sum_i \int_{Q_i+E_{\rm cut}}^{\infty}
\Phi(E_{\nu}) \ \sigma_i(E_{\nu}) \ dE_{\nu}.
\ee
The quantity $E_{\rm cut}=4.489$ MeV is the minimum electron kinetic energy
observable in the ICARUS detector, while the function  $\Phi(E_\nu)$
associated with the $^{\sss 8}$B spectrum was taken from ref.~\cite{15}.
Multiplying $ \Sigma_{\rm tot} $ by the total integrated neutrino flux
${\cal F}(^{\sss 8}{\rm B}) = 5.8 \times 10^6 $ cm$^{-2}$ s$^{-1}$, one
obtains the neutrino capture rate on $^{\sss 40}$Ar
\be
{\cal R}(^{\sss 40}{\rm Ar}) \ = \ \sum_i \ {\cal R}_i(^{\sss 40}{\rm Ar})
\ = \ \Sigma_{\rm tot}(^{\sss 40}{\rm Ar}) \ {\cal F}(^8{\rm B}).
\ee

For the $\beta$-decay of $^{\sss 40}$Ti
\be
^{\sss 40}{\rm Ti} \ \longrightarrow \ ^{\sss 40}{\rm Sc} \ + \ e^+ \ + \
\nu  \, ,
\ee
the partial halflife for the decay to the $i^{th}$ state in $^{\sss 40}$Sc
is given by
\be
t_{1/2}^i = \frac{K}{G_V^2\mid {\cal M}_{o \rightarrow i} \mid^2 f_{o
\rightarrow i} \, (1 + \delta_R)},
\ee
where $K=2\pi^3(\ln 2)\hbar^7/(m_e^5c^4)$, and we use the value
$K/G_V^2=6170\pm 4 $~s~\cite{16}. For the statistical rate function $f_{o
\rightarrow i}$, we use the formalism of ref.~\cite{Wilk}, which is
expected to be accurate to within 0.5\%, namely
\be
f_{o \rightarrow i}=\int_1^{W_0} {\rm d}W \, p W(W_0-W)^2F_0(Z,W)L_0(Z,W)
C(Z,W)R(W)
\ee
where $p$ and $W$ are the electron momentum and energy, respectively, in
units of $m_ec^2$, $W_0$ being the endpoint, and $F_0(Z,W)$, $L_0(Z,W)$,
$C(Z,W)$, and $R(W)$ are parameterized correction factors given in
ref.~\cite{Wilk}. The radiative correction $\delta_R$ is given by
ref.~\cite{Blin}
\be
\delta_R = \frac{\alpha}{2\pi} \frac{\int{\rm d}W \, p W(W_0-W)^2g(W,W_0)}
{\int{\rm d}W \, p W(W_0-W)^2},
\ee
where $g(W,W_0)$ is given by Eq.~(III-21) in ref.~\cite{Blin}. The total
halflife is then given by the sum of decay rates, i.e.
\be
\frac{1}{t_{1/2}} = \sum_i \frac{1}{t_{1/2}^i},
\ee
while the branching ratio to the $i^{th}$ state is given by
\be
{\rm BR}_i = \frac{t_{1/2}}{t_{1/2}^i}.
\ee

Shown in Table~1 are the explicit values for transitions to each of the
$J^\pi=1^+$, $T=1$ levels in $^{\sss 40}$K ($^{\sss 40}$Sc for the
$\beta$-decay of $^{\sss 40}$Ti).  For the excitation energies, theoretical
values are tabulated, as well as the experimental values for the first
seven $J^\pi=1^+$ levels as determined from the $^{\sss 40}$K spectrum. One
sees that there is a one-to-one correspondence between the theoretical and
experimental levels, with the theoretical levels having an excitation
energy of approximately 0.5--1.0~MeV higher than experiment. For the
purpose of computing the neutrino cross sections and the $\beta$-decay
partial half-lives, the experimental energies were used whenever possible.
Also presented in the Table are the theoretical branching ratios for the
$\beta$-decay of $^{\sss 40}$Ti and their experimental values as deduced in
ref.~\cite{8}. Note that the experimental values are lower limits as they
involve the decay process in which the proton-unbound excited state of
$^{\sss 40}$Sc decays directly to the ground state of $^{\sss 39}$Ca. For
completeness, the Table also reports the parameters for the Fermi
transition to the isobaric analog state.

The primary conclusion of the shell-model calculation is that there is
significant low-lying Gamow-Teller strength that leads to an overall
enhancement of the neutrino absorption cross section by about a factor
three over that expected from the Fermi transition alone. This conclusion
is also supported by existing experimental data, where we find good
agreement between the theoretical halflife $ t_{1/2}  = 53$~ms and the
experimental value $\tau_{1/2} = 56^{+18}_{-12}$~ms~\cite{8}. In addition,
the limited branching-ratio data indicates that most of the Gamow-Teller
strength is to the second $J^\pi=1^+$ state ($\geq 20$\%) as is predicted
by theory.

The main drawback of the theoretical calculation, however, is that it is
difficult to predict the magnitude of the uncertainty in the results. This
is evident in Table~1 where the shell-model underpredicts the branching
ratio to the first $J^\pi=1^+$ state. To be noted, however, that even in
complete shell-model calculations (i.e. without truncation and with a more
sophisticated treatment of the quenching factors)  the strengths obtained
for  the weakest states can show large deviations from their experimental
values, while the strengths of the strongest states are usually more
reliable. Aside from these uncertainties and keeping with the fact that one
is trying to obtain reliable numbers to calibrate a high-energy neutrino
detector, the most preferable course would be to determine the B(GT) values
experimentally. Generally speaking, this is not possible since the nucleus
of interest is usually stable. Instead, B(GT) values must be extracted from
(n,p) or (p,n) reaction studies, or from the $\beta$-decay of the mirror
nucleus by exploiting the fact that isospin is a conserved quantity to
within a few percent. In fact, in the absence of the Coulomb potential (as
well as the smaller nuclear charge-dependent interaction) and second-class
weak currents, B(GT) values for $\beta^-$ and $\beta^+$ decays of mirror
nuclei in the same isospin multiplet are identical. For the most part, in
comparing the $ft$ values of mirror analogs, the largest correction that
needs to be applied is the effect due to different binding energies in the
two analogs (cf. ref.~\cite{Wilk} page 980 and ref.~\cite{Blin} page 107),
while the corrections due to second-class currents are probably quite
small. Altogether, the experimental results for even-A nuclei in the range
$ 8 \leq {\rm A} \leq 30 $ give a correction which appears to decrease in
magnitude with increasing A, being  of the order of 2--4\% for the heavier
cases \cite{Wilk}. This is, however, a point that deserves further careful
study since no definite limits have been placed on second-class currents so
far and since the calculation of the  effects due to binding is also an
open problem.

Because of the large Q-value for the $\beta^+$-decay of $^{\sss 40}$Ti, it
is in principle possible to measure all the B(GT) values of interest for
neutrino absorption on $^{\sss 40}$Ar directly. The primary difficulty in
making this measurement is that $^{\sss 40}$Ti $\beta$-decays into levels
of $^{\sss 40}$Sc that are proton unbound. In ref.~\cite{8}, where the
actual goal of the experiment was to observe direct two-proton
radioactivity, only transitions involving proton emission to the ground
state of $^{\sss 39}$Ca were analyzed. As a result, not enough information
exists to deduce experimental B(GT) values. Therefore, we strongly suggest
that the experiment be repeated with a new focus. Namely, the {\em full}
detection of $\beta^+$-transitions in $^{\sss 40}$Ti to $^{\sss 40}$Sc and
their corresponding proton decays to $^{\sss 39}$Ca, so that the branching
ratios relevant to the $\beta^+$-decay of $^{\sss 40}$Ti may be obtained.
This course of action would, in point of fact, be of crucial importance for
the ICARUS experiment in the context of the study of higher energy solar
neutrinos, allowing an accurate calibration of the associated detector. \\

We would like to thank E.~Calligarich, R.~Dolfini, and M.~Terrani of the
ICARUS collaboration for bringing this subject to our attention and
P.~Vogel for many illuminating and useful discussions. WEO acknowledges the
Weingart Foundation for financial support, as well as partial support from
National Science Foundation grants PHY90-13248 and PHY91-15574 and the
Department of Energy grant DE-FG03-88ER40397. WEO also wishes to thank the
Istituto Nazionale di Fisica Nucleare, Sezione di Milano for its kind
hospitality.

\newpage
\bibliographystyle{try}
\bibliography{ar40}
\newpage
\setlength{\baselineskip}{20pt}
{\small
{\bf TABLE 1 -- } Results of the shell-model calculation for the neutrino
capture on $^{\sss 40}$Ar and for the $\beta$-decay of $^{\sss 40}$Ti. The
available experimental values for $E_i$ and BR$_i$, and the parameters for
the Fermi  transition (IAS) are also reported. The different quantities are
explained in the text.
\begin{center}
\begin{tabular}{||c | c |c |c | c | c | c| c | c| c|| }   \hline  \hline &
& & & & & & & & \\  $ \ i \ $ & $  E_i$(th)  &  $  E_i$(ex)  &  $ \mid
{\cal M}_{o \rightarrow i} \mid^2$  & $  f_{o \rightarrow i} $ & $ \
t_{1/2}^i \ $  & BR$_i$(th) & BR$_i$(ex) & $ \ \Sigma_i \ $ & ${\cal
R}_i$($^{\sss 40}$Ar)\\ & {\footnotesize (MeV)} & {\footnotesize (MeV)}& &
& {\footnotesize (sec)}  & {\footnotesize (\%)}& {\footnotesize (\%)} &
{\footnotesize ($ 10^{\scriptscriptstyle -46}$ cm$^{\scriptscriptstyle
2}$)} & {\footnotesize (SNU)}\\ \hline \hline & & & & & & & & &  \\ 1
&2.684 &2.290 &0.006 &35455 &29.0 & 0.18& 4 & 20.21 & 0.012\\  & & & & & &
& & &  \\ 2 &2.971 &2.730 & 1.195&27972 & 0.185 & 28.72 & 20  & 3262.28 &
1.892\\ & & & & & & & & &  \\ 3 & 3.291&3.110 & 0.946&21821 & 0.299 & 17.7&
3 & 2075.09 & 1.204\\ & & & & & & & & &  \\ 4 &3.622 &3.146 & 0.101&
21337&2.86 & 1.85 & -- & 218.54 & 0.127\\ & & & & & & & & &  \\ 5 &
4.308&3.739 & 0.034& 14524& 12.1 &0.44 & -- & 50.68 & 0.029\\ & & & & & & &
& &  \\ 6 & 4.521&3.798 & 1.119&13955 &0.395 & 13.42& -- & 1643.01 &
0.953\\ & & & & & & & & &  \\ {\footnotesize IAS} &-- & 4.384& 4& 9819 &
0.157 & 33.74 &16  & 3847.32 & 2.23\\ & & & & & & & & &  \\ 7 & 4.801&4.789
& 0.082&6761 & 11.1 & 0.48& -- & 57.34 & 0.033\\ & & & & & & & & &  \\ 8 &
5.282&-- & 0.239&4511 &5.72 & 0.93& -- & 106.13  & 0.062\\ & & & & & & & &
&  \\ 9 &5.642 &-- & 0.030&3282 &62.7 & 0.08& -- & 9.12  & 0.005\\ & & & &
& & & & &  \\ 10 &5.823 &-- & 0.000&2774 &$\infty $ & 0& -- & 0  & 0\\ & &
& & & & & & &  \\ 11 &5.922 &-- & 0.383&2524 & 6.38 & 0.83 & -- & 81.35 &
0.047\\ & & & & & & & & &  \\ 12 &6.151 &-- &0.023 &2014 & 133.2 & 0.04&--
& 3.76 & 0.002\\ & & & & & & & & &  \\ 13 & 6.428&-- & 0.698&1510 &5.85 &
0.91 & -- & 76.15 & 0.044\\ & & & & & & & & &  \\ 14 & 6.480&-- &
0.343&1428 & 12.6 & 0.42&--  & 36.75 & 0.021\\ & & & & & & & & &  \\ 15 &
6.683& --& 0.052&1141 &104.0 & 0.05 & -- & 4.11 & 0.002\\ & & & & & & & & &
 \\ 16 &6.876 & --& 0.017&912 & 398.0 & 0.01 & -- & 0.97 & 0.001\\ & & & &
& & & & &  \\ 17 &7.087 & --& 0.121&705 & 72.3 & 0.07& -- & 4.77 & 0.003\\
& & & & & & & & &  \\ 18 &7.123 & --& 0.002&674 & 4577 & 0 &--  & 0.07 &
0\\ & & & & & & & & &  \\ 19 &7.368 & --& 0.233&489 & 54.8 &0.10 &--  &
5.02 & 0.003\\ & & & & & & & & &  \\   \hline   \hline \end{tabular}
\end{center}
}

\end{document}